\documentclass[preprint,
showpacs,amsmath
,nofootinbib
]{revtex4-1}
\usepackage{graphicx}
\usepackage{dcolumn}
\usepackage{bm}
\usepackage{epstopdf, epsf}
\usepackage{hyperref}
\everymath{\displaystyle}
\begin{document}
\title{Notes on a scalar field with a kinetic term non-minimally coupled to gravity}
\author{Amir Ghalee}
\email[\,Electronic address:\ ]{ghalee@ut.ac.ir}
\affiliation{{Department of Physics, Tafresh  University,
P. O. Box 39518-79611, Tafresh, Iran}}
\begin{abstract}
We explore the inflationary phase of a scalar field with a kinetic term non-minimally coupled to gravity. We find that one
of the slow-roll conditions is naturally consequence of the equation of motion of the scalar field. Thus, slow-roll conditions impose fewer constraints on potentials than other inflationary models. Moreover, it is demonstrated that the inflationary phase can be described by just one slow-roll parameter. By investigating the metric perturbations, it is shown that except for one potential, almost all potentials have the same pattern in the ($n_{s}$, $r$) plane. We provide an exact solution for the exceptional case. The exact solution represents the condensed scalar field and results in an accelerated expansion.
\end{abstract}
\maketitle
\section{\label{sec:level1}INTRODUCTION}
Observation of the cosmic microwave background and large scale structure are consistent with slow-roll inflation paradigm in which
a scalar field rolls slowly down its potential \cite{plank-data}.\\
To have slow-roll inflationary phase, certain slow-roll conditions are considered that lead to some constraints
on parameters of a model \cite{weinberg}. For example,
consider a simple theory of inflation described by the following Lagrangian
\begin{equation}\label{simple}
\mathcal{L_{s}}=\sqrt{-g}\left[\frac{R}{2\kappa^{2}}-\frac{1}{2}g^{\mu\nu}\partial_{\mu}\phi\partial_{\nu}\phi-V(\phi)\right],
\end{equation}
where $\kappa^{2}=8\pi G$, in the flat Friedmann-Robertson-Walker metric with signature $(-,+,+,+)$ and $a$ as the scale factor. There exist two \emph{independent} slow-roll conditions. The first of these conditions is actually the definition of inflationary phase states that $\dot{H}/H^{2}\equiv\epsilon\ll1$.
Since the evolution of the scalar field is given by a second-order differential equation, it is not clear that the first condition holds over an extended period. So, the second condition is required in which we \emph{demand} that $\ddot{\phi}\ll H\dot{\phi}$.\\
Also, to determine the details of models from observations, various slow-roll parameters are used such as $\epsilon, \eta\equiv\dot{\epsilon}/H\epsilon, \chi\equiv\dot{\eta}/H\eta $. The parameters are used to determined the specific potential term in \eqref{simple} for the inflation era \cite{plank-data}. Since the shape of the potential term is important to study high energy physics,
 the values of these parameters connect cosmology to particle physics.\\
In this paper we introduce a model for which we need just one parameter, $\epsilon$, to describe slow-roll inflation phase. The model is
given by the following action
\begin{equation}\label{0-1}
S=\int d^{4}x\sqrt{-g}\left[\frac{R}{2\kappa^{2}}-g^{\mu\nu}\partial_{\mu}\varphi\partial_{\nu}\varphi++\frac{1}{2}\alpha^{2}G^{\mu\nu}\partial_{\mu}\varphi\partial_{\nu}\varphi-V(\varphi)
\right],
\end{equation}
where $G^{\mu\nu}$ is the Einstein's tensor and $\alpha$ is an inverse mass parameter.
 We have chosen $+$ sign for the second term because
for the de Sitter space we have $G_{\mu\nu}\propto-g_{\mu\nu}$, so the scalar filed in \eqref{0-1} has the same dynamics as the scalar filed in \eqref{simple} in the de Sitter space.\\
The third term is one of the operators of the Horndeski's scalar-tensor theory \cite{Horndeski}. Thus, in this model we confront with second-order differential equations.
C. Germani et al studied $V(\varphi)=\lambda\varphi^4$ for "the new higgs inflation" \cite{higgs,higgse} and extended their model for other cases in \cite{higgs-extend}. It has been shown that a new type of inflationary phase arises when the third term dominates over the standard kinetic term \cite{higgs,higgs-extend}. So in this paper we will focus on this type of inflation and we will neglect the second term in the action \eqref{0-1}.\\
Compared to the other works \cite{higgs,higgs-extend}, we have two new results. The first result is that for $V(\varphi)=M^{6}/\varphi^2$ background equations can be solve exactly and metric perturbations will be obtained. The second result is to show that up to the first order in $\epsilon$, in contrast to the simple model of inflation \eqref{simple}, the values of measurable
quantities of the inflationary regime of this model are almost independent from the form of the potential term in \eqref{0-1}, as is shown in Fig. 1. Also, the model gives values for the parameters which are agreement with the Planck data in \cite{plank-data}.\\
The organization of this paper is as follows: in Sec. II an exact solution for a special potential is given and the metric perturbations of the solution is discussed. we show that the scalar field is condensed and results in accelerated expansion.
Sec. III is devoted to study the background and the metric perturbations of general potentials. We show that many features of the exact solution are similar to other potentials. We summarize our findings and discuss about the results in Sec. IV. In Appendix the
second order actions for the metric perturbations, in which $V(\varphi)\neq0$, is given.
\section{\label{sec:level1} Exact Solution}
In this section we consider $V(\varphi)=M^{6}/\varphi^{2}$ and the exact solution for the model is provided, which is also the attractor solution.
Many properties of the solution, are similar to other potentials. So, it can be used to check many
results and approximations in the next section. As we will see, the unique feature of this case is that inflation would
not end \cite{weinberg}.\\
Let us show that the following solution is the exact and the attractor solution of the model for the stated potential:
\begin{equation}\label{claim1}
\varphi=M^{2}t,\hspace{.5cm} H=\frac{1}{\epsilon\hspace{.04cm} t},
\end{equation}
where $M$ is a constant with dimension of mass which is determined by
\begin{equation}\label{claim2}
\alpha^{2}\kappa^{2} M^{4}=\frac{2\epsilon}{3+\epsilon},
\end{equation}
and $\epsilon<1$. Thus, the scalar field is condensed and results in an accelerated expansion phase.\\
We first prove the above statements and then investigate the metric perturbations for the solution.
\subsection{\label{sec:level1} Background cosmology}
The Friedmann equation and the equation of motion for the scalar field are obtained as \cite{higgs}
\begin{subequations}
\begin{align}
        &H^{2}=\frac{\kappa^{2}}{6}\left[\dot{\varphi}^{2}9\alpha^{2}H^{2}+2V(\varphi)\right]\label{1-1-1},\\
       &\ddot{\varphi}-3H\dot{\varphi}w_{eff}=-\frac{1}{3H^{2}\alpha^{2}}\frac{dV(\varphi)}{d\varphi}\label{1-1-2},
\end{align}
\end{subequations}
where $w_{eff}$ is the effective equation of state
\begin{equation}\label{1-2}
w_{eff}=-1-\frac{2}{3}\frac{\dot{H}}{H^{2}}\quad.
\end{equation}
For $V(\varphi)=M^{6}/\varphi^{2}$, by inserting \eqref{claim1} in \eqref{1-1-1} it follows that
\begin{equation}\label{1-3}
\kappa^{2}M^{2}=\frac{3(3-2\epsilon)}{\epsilon^{2}(\epsilon+3)}.
\end{equation}
So, at this step, we have an upper bound $\epsilon<3/2$ which results in $w_{eff}<0$.\\
Inserting \eqref{claim1} in \eqref{1-1-2} yields
\begin{equation}\label{1-4}
\alpha^{2}M^{2}=\frac{2\epsilon^{3}}{3(3-2\epsilon)}.
\end{equation}
Therefore, Eq. \eqref{claim2} follows from Eqs. \eqref{1-3} and \eqref{1-4}.\\
Just for a comparison between this case and other potentials in the next section; by combining expressions in \eqref{claim2}, \eqref{1-4} and
taking $\epsilon\ll1$, it turns out that
\begin{equation}\label{compa}
\alpha M\ll\alpha^{2}\kappa^{2}M^{4}\ll1
\end{equation}
To show that the solution is attractor solution, consider a small perturbation $\delta\varphi$ as $\varphi=M^{2}t+\delta\varphi$.
Using \eqref{1-1-2} and the following relation
\begin{equation}\label{1-5}
w_{eff}=-1-\frac{2}{3}\epsilon,
\end{equation}
it turns out that
\begin{equation}\label{1-6}
\delta\ddot{\varphi}-\frac{3}{\epsilon\hspace{.05cm}t}w_{eff}\delta\dot{\varphi}-\frac{9}{\epsilon\hspace{.05cm} t^{2}}w_{eff}\delta\varphi=0.
\end{equation}
So, $\delta\varphi$ has power law solution as $\delta\varphi\propto t^{\xi}$, where $\xi$ satisfy in the following equation
\begin{equation}\label{1-7}
\xi^{2}-\left(\frac{3}{\epsilon}w_{eff}+1\right)\xi-\frac{9}{\epsilon}w_{eff}=0.
\end{equation}
We pointed out that $w_{eff}<0$, so if we demand that
\begin{equation}\label{1-8}
\frac{3}{\epsilon}w_{eff}+1<0
\end{equation}
The both solutions for $k$ have negative real part an then both solutions for $\delta\varphi$ decay for increasing time, while $\varphi$ itself is increasing.\\
Using Eq. \eqref{1-5} and the above condition, we have
\begin{equation}\label{1-9}
\epsilon<1.
\end{equation}
Therefore, by just demand reasonable conditions for our solution, the accelerated expansion is emerged.
\subsection{\label{sec:level1}Metric perturbations}
The second order action for the metric perturbations are given in
\cite{higgs-extend} and  we quote the results in the appendix.\\
 So far,we have shown that to have inflationary era for $V(\varphi)=M^{6}/\varphi^{2}$, it is not necessary to demand slow roll approximations. As we will see,this aspect of the model gives different predictions for this potential with respect to the other potentials \cite{weinberg}.\\
We use comoving gauge in which the scalar metric perturbations, $\zeta$, and the tensor metric perturbations ,$\gamma_{ij}$, are defined as
\begin{equation}\label{1-10}
  \delta\varphi=0\quad h_{ij}=a^{2}[(1+2\zeta)\delta_{ij}+\gamma_{ij}],
\end{equation}
where $\partial_{i}\gamma_{ij}=0$, $\gamma^{ii}=0$.
For the scalar metric perturbation, from Eqs. \eqref{claim1} and \eqref{Aaction} it follows that
\begin{equation}\label{1-11}
 S^{s}=\frac{1}{\kappa^{2}}\int d^{4}xD_{s}^{2}\left[c_{s}^{2}a\zeta\partial^{2}\zeta+a^{3}\dot{\zeta}^{2}\right].
\end{equation}
Where
\begin{equation}\label{1-12}
 D_{s}^{2}\equiv\frac{\epsilon(4\epsilon+3)(1-\frac{2}{3}\epsilon)^{2}}{(3+\epsilon)}(1+\frac{2}{3}\epsilon)^{\frac{3}{2}},
\end{equation}
and the speed of sound is
\begin{equation}\label{1-13}
c_{s}^{2}\equiv\frac{(9+4\epsilon)(3-2\epsilon)}{3(3+2\epsilon)(3+4\epsilon)}.
\end{equation}
To obtain \eqref{1-11}, we also re-scaled the scale factor as
\begin{equation}\label{trans}
a\rightarrow\sqrt{1+\frac{2}{3}\epsilon}\hspace{.15cm}a.
\end{equation}
When we want to study the tensor metric perturbation, the reason for the above re-scaling will be clarified. Of course we must
it for the background quantities and the tensor metric perturbation. Here $\epsilon$ is just a number, so the re-scaling
$H\rightarrow H$.\\
Also, note that since $0<\epsilon<1$, from the above expression it turns out that $0<c_{s}^{2}<1$.\\
Here $D_{s}$ and $c_{s}^{2}$ are just numbers. We represent them to compare them with our results in the next section, in which the calculations
are done to first order in $\epsilon$. For $\epsilon\ll1$ we have
\begin{equation}\label{1-14}
 D_{s}^{2}=\epsilon+\mathcal{O}(\epsilon^{2}),
\end{equation}
and
\begin{equation}\label{1-15}
c_{s}^{2}=1-\frac{20}{9}\epsilon+\mathcal{O}(\epsilon^{2}).
\end{equation}
From \eqref{1-11}, by variation with respect to $\zeta$ and using the Fourier modes of $\zeta$ as
\begin{equation}\label{1-16}
\zeta=\int\frac{d^{3}k}{(2\pi)^{\frac{3}{2}}}\zeta_{k}e^{i\vec{k}.\vec{x}},
\end{equation}
it turn out that
\begin{equation}\label{1-17}
\ddot{\zeta}_{k}+3H\dot{\zeta}_{k}+(\frac{c_{s}k}{a})^{2}\zeta_{k}=0.
\end{equation}
Using conformal time, $\tau=\int \frac{dt}{a}$, Eq. \eqref{1-17} becomes
\begin{equation}\label{1-18}
\zeta''_{k}-\frac{2}{(1-\epsilon)\tau}\zeta'_{k}+(c_{s}k)^{2}\zeta_{k}=0.
\end{equation}
Where $'$ denotes derivative with respect to $\tau$.\\
The above equation has two solutions as $(-\tau)^{\nu}H_{\nu}^{(1)}(-c_{s}k\tau)$ and $(-\tau)^{\nu}H_{\nu}^{(2)}(-c_{s}k\tau)$.
Where $H_{\nu}^{(1)}(x)$ and $H_{\nu}^{(2)}(x)$ are the Hankel functions, for which $H_{\nu}^{(1)}(x)=H_{\nu}^{(2)*}(x)$, and
\begin{equation}\label{1-19}
\nu=\frac{3}{2}+\frac{\epsilon}{1-\epsilon}.
\end{equation}
From asymptotic behavior of the Hankel function, the inside the horizon limit, $-c_{s}k\tau\gg1$, is given by
\begin{equation}\label{1-20}
\zeta_{k}\rightarrow(-c_{s}k\tau)^{\nu-\frac{1}{2}}(c_{s}k)^{-\nu}\exp[i(-c_{s}k\tau-\nu\frac{\pi}{4}-\frac{\pi}{4})].
\end{equation}
Here, the Bunch-Davies vacuum is adopted and we take $H_{\nu}^{(1)}(x)$ as the solution.\\
Also, the outside the horizon limit, $-c_{s}k\tau\ll1$, is given by
\begin{equation}\label{1-21}
\zeta_{k}\rightarrow (c_{s}k)^{-\nu}.
\end{equation}
Therefore, the power spectrum can be obtained from
\begin{equation}\label{1-21-1}
\left.<\zeta_{k}\zeta_{k'}>\right|_{kc_{s}\tau=-1}=\frac{2\pi^{2}}{k^{3}}P_{s}\delta^{3}(k+k'),
\end{equation}
which results in
\begin{equation}\label{1-22}
P_{s}=A_{s}\frac{c_{s}^{-2\nu}}{2\pi^{2}}k^{3-2\nu},
\end{equation}
where $A_{s}$ is the scalar amplitude.\\
Thus, the spectral index of scalar perturbation is obtained as
\begin{equation}\label{1-23}
n_{s}=4-2\nu.
\end{equation}
As for the tensor metric perturbation, $\gamma_{ij}$, by using \eqref{claim1} and \eqref{Action2} we have
\begin{equation}\label{1-23-1}
S^{T}=\frac{3}{8(\epsilon+3)\kappa^{2}}\int d^{4}x\left[(1+\frac{2}{3}\epsilon)a\gamma_{ij}\partial^{2}\gamma_{ij}+a^{3}\dot{\gamma_{ij}}^{2}\right].
\end{equation}
Now, to absorb $(1+\frac{2}{3}\epsilon)$ into the scale factor, we can use \eqref{trans}. We pointed out that we must apply it for the background quantities and
the tensor metric perturbation, as we did it.
So, using \eqref{trans} it turns out that
\begin{equation}\label{1-25}
 S^{T}=\frac{D^{2}_{t}}{8\kappa^{2}}\int d^{4}x\left[a\gamma_{ij}\partial^{2}\gamma_{ij}+a^{3}\dot{\gamma_{ij}}^{2}\right],
\end{equation}
where
\begin{equation}\label{tadd}
D^{2}_{t}\equiv\frac{(1+\frac{2}{3}\epsilon)^{\frac{3}{2}}}{1+\frac{\epsilon}{3}}.
\end{equation}
So, again just for calculations in the next section, by taking $\epsilon\ll1$ results in
\begin{equation}\label{tadd}
D^{2}_{t}=1+\mathcal{O}(\epsilon).
\end{equation}
From \eqref{1-25}, variation with respect to $\gamma_{ij}$ and using the following Fourier representation
\begin{equation}\label{1-26}
 \gamma_{ij}=\int\frac{d^{3}k}{(2\pi)^{3/2}}\sum_{s=\pm}\epsilon_{ij}^{s}(k)\gamma_{k}^{s}(t)e^{i\overrightarrow{k}.\overrightarrow{x}},
\end{equation}
 where $\epsilon_{ii}=k^{i}\epsilon_{ij}=0$ and $\epsilon_{ij}^s(k)\epsilon_{ij}^{{s'}}(k)=2\delta_{s'}$, leads to
 the following equation for the two polarization modes of the gravitation waves($+$ and $-$)
 \begin{equation}\label{1-27}
\ddot{\gamma_{k}^{s}}+3H\dot{\gamma_{k}^{s}}+(\frac{k}{a})^{2}\gamma_{k}^{s}=0.
\end{equation}
Thus, in terms of the conformal time, $\tau$, we have
\begin{equation}\label{1-28}
\gamma_{k}^{s''}-\frac{2}{(1-\epsilon)\tau}\gamma_{k}^{s'}+k^{2}\gamma_{k}^{s}=0.
\end{equation}
The above equation is similar to \eqref{1-18}, and inside the horizon limit, $-k\tau\gg1$, is obtained as
\begin{equation}\label{1-29}
\gamma_{k}^{s}\rightarrow(-k\tau)^{\mu-\frac{1}{2}}(k)^{-\mu}\exp[i(-k\tau-\mu\frac{\pi}{4}-\frac{\pi}{4})],
\end{equation}
where
\begin{equation}\label{1-mr}
\mu=\frac{3}{2}+\frac{\epsilon}{1-\epsilon}.
\end{equation}
The outside the horizon limit, $-k\tau\ll1$, is given by
\begin{equation}\label{1-30}
\gamma_{k}^{s}\rightarrow (k)^{-\mu}.
\end{equation}
Therefore, the tensor power spectrum for each polarization mode of $\gamma_{ij}$ can be obtained from
\begin{equation}\label{1-21-1}
\left.<\gamma_{k}\gamma_{k'}>\right|_{kc_{s}\tau=-1}=\frac{2\pi^{2}}{k^{3}}P_{s}\delta^{3}(k+k'),
\end{equation}
which results in
\begin{equation}\label{1-31}
P_{t}=\frac{A_{t}}{\pi^{2}}k^{3-2\mu},
\end{equation}
where $A_{t}$ is the tensor amplitude.\\
From \eqref{1-22}, and \eqref{1-31} the tensor-to-scalar power ratio, $r$, is given by
\begin{equation}\label{1-32}
r=\frac{A_{t}}{A_{s}}c_{s}^{2\nu}.
\end{equation}
At this step $A_{s}$ and $A_{t}$ are arbitrary constants. To normalized $r$, note that if $\epsilon\rightarrow0$, i.e. de Sitter limit, then $c_{s}\rightarrow1$. But in de Sitter space $G_{\mu\nu}\propto-g_{\mu\nu}$, and the action, \eqref{0-1}, takes similar form as \eqref{simple}. Therefore,
 we can use the usual standard normalization which is
\begin{equation}\label{1-33}
\frac{A_{t}}{A_{s}}=16\epsilon,
\end{equation}
Thus
\begin{equation}\label{1-34}
r=16c_{s}^{2\nu}\epsilon.
\end{equation}
If we take the following values for $n_{s}$
\begin{equation}\label{exf1}
    0.95<n_{s}<0.98\hspace{.1cm},
\end{equation}
from the above relations, we give
\begin{equation}\label{exf}
 0.15<r<0.37\hspace{.1cm},\hspace{.15cm} 0.947<c_{s}^{2}<0.978\hspace{.08cm}.
\end{equation}
As is shown by the dashed line in Figure 1. By comparing with the Plank data in \cite{plank-data}, it turns out that this potential is not suitable for the inflation era in the early Universe. 
\section{\label{sec:level1}Other potentials}
In this section we prove the following statements and investigate consequences of them.\\
Apart from $V(\varphi)=M^{6}/\varphi^{2}$, for any potential in \eqref{0-1} the sufficient conditions to have inflationary phase are
\begin{equation}\label{claim3}
\left\lvert\frac{\partial V(\varphi)}{\partial\varphi}\right\rvert\ll\alpha\kappa^{2}V^{3/2}(\varphi),\hspace{1cm}\alpha\ll\kappa\hspace{.02cm} V(\varphi).
\end{equation}
Note that the first condition in the above expression is consistent with \eqref{compa}. We will show that why the second condition can not be applied for the exceptional case.\\
Also, for any potential in the inflationary phase we have
\begin{equation}\label{claim4}
\frac{3}{2}\alpha^{2}\kappa^{2}\dot{\varphi}^{2}\simeq\epsilon\ll1,
\end{equation}
which is also consistent with \eqref{compa}. Thus, although in general case the scalar field is not condensed, it evolves very slowly during the inflationary phase.\\
\subsection{\label{sec:level1}Background cosmology}
Consider situation in which the potential term in \eqref{1-1-1} is dominated, i.e.
\begin{equation}\label{preassh}
\frac{3}{2}\kappa^{2}\alpha^{2}\dot{\varphi}^{2}\ll1,
\end{equation}
so, it follows that
\begin{equation}\label{2-1}
H^{2}\simeq\frac{\kappa^{2}}{3}V(\varphi).
\end{equation}
Hence
\begin{equation}\label{2-2t}
-\frac{\dot{H}}{H^{2}}\simeq-\frac{\kappa^{2}}{6}\frac{\dot{V}(\varphi)}{H^{3}}.
\end{equation}
Multiplying Eq. (1.1.2) with $\kappa^{2}\alpha^{2}\dot{\varphi }/H$, and using $\dot{\varphi}\frac{d}{d\varphi}=\frac{d}{dt}$, then from \eqref{2-2t} results in
\begin{equation}\label{2-3}
\frac{3}{2}\kappa^{2}\alpha^{2}\dot{\varphi}^{2}\left(\frac{\ddot{\varphi}}{3H\dot{\varphi}}+1\right)=-\frac{\dot{H}}{H^{2}}\left(1+\kappa^{2}\alpha^{2}\dot{\varphi}^{2}\right).
\end{equation}
Note that the above result is obtained without assuming inflation period.\\
Now, we first prove that during inflation era we have
\begin{equation}\label{proas}
\frac{3}{2}\kappa^{2}\alpha^{2}\dot{\varphi}^{2}\simeq\epsilon\ll1.
\end{equation}
To obtain the above result, by imposing the slow-roll condition, $\ddot{\varphi}\ll H\dot{\varphi}$, in \eqref{2-3} we have
\begin{equation}\label{proas1}
\frac{3}{2}\kappa^{2}\alpha^{2}\dot{\varphi}^{2}\simeq\epsilon\left(1+\kappa^{2}\alpha^{2}\dot{\varphi}^{2}\right),
\end{equation}
so
\begin{equation}\label{proas2}
\frac{3}{2}\kappa^{2}\alpha^{2}\dot{\varphi}^{2}(1-\frac{2}{3}\epsilon)\simeq\epsilon.
\end{equation}
Thus, for the inflation period, $\epsilon\ll1$, we have
\begin{equation}\label{proas3}
\frac{3}{2}\kappa^{2}\alpha^{2}\dot{\varphi}^{2}\simeq\epsilon\left(1+\mathcal{O}(\epsilon)\right).
\end{equation}
Note that the above result is consistent with \eqref{preassh}.\\
Returning now to Eq. \eqref{2-3}, by taking Eq. \eqref{proas} it follows
\begin{equation}\label{2-4}
\frac{\ddot{\varphi}}{H\dot{\varphi}}\simeq2\epsilon\ll1. \hspace{.1cm} (for\hspace{.2cm} \dot{\epsilon}\neq0)
\end{equation}
For $\dot{\epsilon}=0$, from Eq. \eqref{proas3} it turns out that $\ddot{\varphi}=0$, which is consistent with the results in the last section.\\
The important point is that one can chose Eq. \eqref{proas} as one of the slow-roll inflation condition, which is actually done by \eqref{preassh}, then Eq. \eqref{2-4} follows from Eq. \eqref{2-3}.\\
To clarify the sharp distinction between this model for inflation and other models, recall that to have a
slow-roll inflation the following slow parameters are defined
\begin{equation}\label{slowparameters}
\epsilon,\hspace{.02cm}\eta\equiv\frac{\dot{\epsilon}}{H\epsilon},\hspace{.02cm} \chi\equiv\frac{\dot{\eta}}{H\eta},\cdots.
\end{equation}
The above quantity should remain small during slow-roll inflation. They are defined to \emph{demand} suppression of the dynamics of a scalar field(s), which is usually caused by second (or higher)-order differential in equations.\\
But in this model, From \eqref{proas} and \eqref{2-4} it turns out that (to first order in $\epsilon$)
\begin{equation}\label{p-3}
\eta\equiv\frac{\dot{\epsilon}}{H\epsilon}\simeq\frac{2\ddot{\varphi}}{H\dot{\varphi}}\simeq4\epsilon.\hspace{.1cm} (for\hspace{.2cm} \dot{\epsilon}\neq0)
\end{equation}
Therefore, \emph{we need just one parameter to describe the inflationary phase of the model}.\\
Again, recall that usually the slow-roll parameters contain fingerprint of different potentials. Thus,
from the above statement, it seems that in this model various potentials have similar behavior in the inflationary phase.
We will show that this expectation is correct.\\
The number of $e$-folding can be read from \eqref{p-3} as
\begin{equation}\label{fold}
\mathcal{N}\equiv\int_{t_{i}}^{t_{f}}Hdt\simeq\frac{1}{4}\int_{t_{i}}^{t_{f}}\frac{\dot{\epsilon}}{\epsilon^{2}}\simeq\frac{1}{4\epsilon(t_{i})}.
\end{equation}
Where, for the end of inflation we take $\epsilon(t_{f})\approx 0.1$.\\
To represent inflationary conditions by potential, from \eqref{2-2t} and \eqref{proas}, it follows that
\begin{equation}\label{con1}
\sqrt{\epsilon}\simeq\frac{1}{\alpha\kappa^{2}} \left\lvert\frac{1}{V^{3/2}(\varphi)}\frac{\partial V(\varphi)}{\partial\varphi}\right\rvert.
\end{equation}
For $\epsilon\ll1$, the above result is agreement with \eqref{claim3}. Note that to obtain the above condition, we used the equations which are
correct for every $\epsilon$, so it can be applied for $V(\varphi)=M^{6}/\varphi^{2}$.\\
Using the definition of $\epsilon$ and Eq. \eqref{2-4}, during inflationary phase, we have
\begin{equation}\label{2-4-solve}
\dot{\varphi}=\frac{1}{H^{2}}.\hspace{.1cm} (for\hspace{.2cm} \dot{\epsilon}\neq0)
\end{equation}
Hence, the condition \eqref{proas} can be read as
\begin{equation}\label{claim3read}
\alpha\hspace{.02cm}\kappa\ll H^{2}\simeq\frac{\kappa^{2}}{3}V(\varphi),\hspace{.1cm}(for\hspace{.2cm} \dot{\epsilon}\neq0)
\end{equation}
which is in agreement with \eqref{claim3}. Here we used the equation which are not correct for $\dot{\epsilon}=0$, so the above condition can not be used for
$V(\varphi)=M^{6}/\varphi^{2}$.\\
By inserting \eqref{2-4-solve} in Eq. \eqref{proas}, the Hubble parameter takes the following form during inflationary phase
\begin{equation}\label{Hubble}
H^{3}=H_{f}^{3}-\frac{9}{2}\alpha^{2}\kappa^{2}t, \hspace{.1cm} (for\hspace{.2cm} \dot{\epsilon}\neq0)
\end{equation}
where $H_{f}$ is a constant of integration.\\
Finally, the field configuration for $V(\varphi)$, can be obtained by \eqref{2-1} and \eqref{Hubble}.\\
As for graceful exit, From \eqref{p-3} it turns out that
\begin{equation}\label{2-5}
\dot{\epsilon}\simeq4H\epsilon^{2}.
\end{equation}
Thus, $\epsilon$ is increased with time, although gradually, and the first term in Eq. \eqref{1-1-1} will be dominated. We pointed out that when $V(\varphi)=0$ the scalar field is condensed and the Universe behaves like a dust matter dominated universe \cite{higgs}. To show that this point is correct for general potentials, assuming $\kappa^{2}V(\varphi)\ll3H^{2}$ in \eqref{1-1-1} that results in $\frac{3}{2}\alpha^{2}\kappa^{2}\dot{\varphi}^{2}\simeq1$. So, by taking derivative with respect to time from \eqref{1-1-1}, it follows that
\begin{equation}\label{gra1}
\dot{V}(\varphi)\simeq2\frac{\dot{H}}{H}V(\varphi).
\end{equation}
Multiplying Eq. \eqref{1-1-2} with $\kappa^{2}\dot{\varphi}$ and using the above results yields
\begin{equation}\label{gra2}
w_{eff}=-1-\frac{2}{3}\frac{\dot{H}}{H^{2}}\simeq\frac{\dot{H}}{H^{2}}\frac{\kappa^2V(\varphi)}{3H^{2}}.
\end{equation}
Thus, by using the stated assumption for this case, we lead to $w_{eff}\approx0$.\\
For $V(\varphi)=M^{6}/\varphi^{2}$, the assumption yields
\begin{equation}\label{gra3}
\kappa^{2}M^{2}\ll\frac{3}{\epsilon^{2}},
\end{equation}
which is inconsistent with \eqref{1-3}. This result means that, for this case, the first term in Eq. \eqref{1-1-1} can not be dominated and
inflation would not end, which is consistent with our statements in the previous section.
\subsection{\label{sec:level1}Metric perturbations}
We have shown that, except for $V(\varphi)=M^{6}/\varphi^{2}$, to have inflationary era we must demand slow-roll approximation. As we pointed out, this case studied in \cite{higgs-extend}. Here we want to show that how the results of \cite{higgs-extend} for other potentials can be compared with the exact result.
. As we will see the differences between results arise when we taking into account the dependence of $\epsilon(t)$ on time in the general case.\\
From Eqs. \eqref{claim4}, Eq. \eqref{Aaction} and using the fact that $\epsilon(t)\ll1$  we have
\begin{equation}\label{p-5}
S^{S}=\frac{1}{\kappa^{2}}\int d^{4}x\hspace{.05cm}\epsilon(t)\left[c_{s}^{2}a\zeta\partial^{2}\zeta+a^{3}\dot{\zeta}^{2}\right].
\end{equation}
Where
\begin{equation}\label{p-6}
c_{s}^{2}=1-\frac{20}{9}\epsilon(t).
\end{equation}
Note that the above results is almost similar to expressions in \eqref{1-11}, with \eqref{1-14} and \eqref{1-15}. This point is not amazing, because to obtain
the above results the higher-order terms, $\mathcal{O}(\epsilon(t)^{2})$, have been neglected.\\
Note that, for the same reason in the previous section, we applied \eqref{trans}. For the background case, the re-scaling leads to
\begin{equation}\label{p-4}
H\rightarrow H\left(1+\mathcal{O}(\epsilon(t))\right),
\end{equation}
where Eq. \eqref{p-3} is used.
So, the effect of \eqref{trans} at the background level can be neglected.\\
Variation with respect to $\zeta$ from \eqref{p-5} , and using \eqref{1-16} results in
\begin{equation}\label{p-7}
\ddot{\zeta}+3H\left(1+\frac{4}{3}\epsilon(t)\right)\dot{\zeta}+\frac{c_{s}^{2}k^{2}}{a^{2}}\zeta=0,
\end{equation}
which is very similar to Eq. \eqref{1-17} with the extra term $\frac{4}{3}\epsilon(t)$, that comes from $\dot{\epsilon}(t)$ and then using \eqref{p-3}.\\
In terms of the conformal time, $\tau$, the above equation takes the following form
\begin{equation}\label{p-8}
\zeta''+2Ha(1+2\epsilon(t))\zeta'+c_{s}^{2}k^{2}\zeta=0,
\end{equation}
where
\begin{equation}\label{p-9}
\frac{d}{d\tau}\left(\frac{1}{Ha}\right)=\epsilon(t)-1.
\end{equation}
To solve \eqref{p-9} the slow-roll approximation can be used \cite{weinberg}. So, to the first order in $\epsilon(t)$ we have
\begin{equation}\label{p-10}
Ha=-\frac{(1+\epsilon(t))}{\tau}.
\end{equation}
By inserting this expresion in Eq.\eqref{p-8}, to the firs order in $\epsilon(t)$, we find
\begin{equation}\label{p-11}
\zeta''-\frac{2(1+3\epsilon(t))}{\tau}\zeta'+c_{s}^{2}k^{2}\zeta=0.
\end{equation}
Again, the two solutions are $(-\tau)^{\nu}H_{\nu}^{(1)}(-c_{s}k\tau)$ and $(-\tau)^{\nu}H_{\nu}^{(2)}(-c_{s}k\tau)$ but now
\begin{equation}\label{p-12}
\nu=\frac{3}{2}+3\epsilon(t).
\end{equation}
Thus, using \eqref{1-22} and \eqref{p-12}, the spectral tilt is
\begin{equation}\label{p-23}
n_{s}=1-6\epsilon(t).
\end{equation}
Note that the above result is correct for almost all potential for which, $\dot{\epsilon}(t)\neq0$, at the first order in $\epsilon(t)$.\\
As for the the tensor metric perturbation, from \eqref{Action2} and by taking into account \eqref{trans} to leading order we find
\begin{equation}\label{p-24}
 S^{T}=\frac{1}{8\kappa^{2}}\int d^{4}x\left[a\gamma_{ij}\partial^{2}\gamma_{ij}+a^{3}\dot{\gamma_{ij}}^{2}\right].
\end{equation}
The above action is the same as \eqref{1-25} for $\epsilon(t)\ll1$. Thus, we have
\begin{equation}\label{p-25}
P_{t}=\frac{A_{t}}{\pi^{2}}k^{3-2\mu},
\end{equation}
where
\begin{equation}\label{p-26}
\mu=\frac{3}{2}+\frac{\epsilon(t)}{1-\epsilon(t)}.
\end{equation}
Finally, using the standard normalization \eqref{1-33}, the tensor-to-scalar power ratio becomes
\begin{equation}\label{p-27}
r=16c_{s}^{2\nu}\epsilon(t).
\end{equation}
Here, $\nu$ is given by \eqref{p-12}.\\
Hence, if we take the following values for $n_{s}$
\begin{equation}\label{f-2}
0.95<n_{s}<0.98,
\end{equation}
it follows that
\begin{equation}\label{f-1}
 0.052 <r<0.129\hspace{.08cm},\hspace{.25cm} 0.981<c_{s}^{2}<0.992\hspace{.08cm}.
\end{equation}
Also, from \eqref{fold}, the stated values for $n_{s}$ give
\begin{equation}\label{foldf}
  30<\mathcal{N}<75.
\end{equation}
As a specific example, we have
\begin{equation}\label{f}
 n_{s}=0.96,\hspace{.1cm} r\approx0.1,\hspace{.08cm}c_{s}^{2}\approx0.985\hspace{.08cm},\hspace{.08cm}\mathcal{N}\approx38.
\end{equation}
The ( $n_{s}$, $r$) plane of the model is shown in Figure 1. By comparing with the Plank data, it turns out that with other potentials the predictions of our model are consistent with observation data.\\
To compare with other works in Ref. \cite{higgs,higgs-extend}, let us consider $V(\varphi)=\frac{1}{2}m^2\varphi^2$. This case studied in section 7.2 of Ref. \cite{higgse}. In this case the author obtained
\begin{equation}\
  n_{s}=0.97, r\simeq 0.08 ,\mathcal{N}=50(Ref. [5]\hspace{.05cm} for V(\varphi)=\frac{1}{2}m^2\varphi^2).
\end{equation}
which is completely agreement with the our results.\\
For $V(\varphi)=\frac{\lambda}{4}\varphi^4$ in Ref. \cite{higgs}, the authors obtained
\begin{equation}
  n_{s}=0.97, r\simeq 0.1 ,\mathcal{N}=56( from\hspace{.08cm} Ref. [4] ).
\end{equation}
But we have
\begin{equation}\
  n_{s}=0.97, r\simeq 0.08.
\end{equation}
To find the source of small deviation from our prediction, note that in Ref. \cite{higgs}, the authors obtained $n_{s}=1-5\epsilon$, which follows from $\frac{\ddot{\varphi}}{H\dot{\varphi}}=\frac{3}{2}\epsilon(t)$ in Ref. \cite{higgs}. But we have obtained $n_{s}=1-6\epsilon$ which follows from
Eq. \eqref{p-3}. Note that we have proved Eq. \eqref{p-3} in general case.
\begin{figure}
\includegraphics{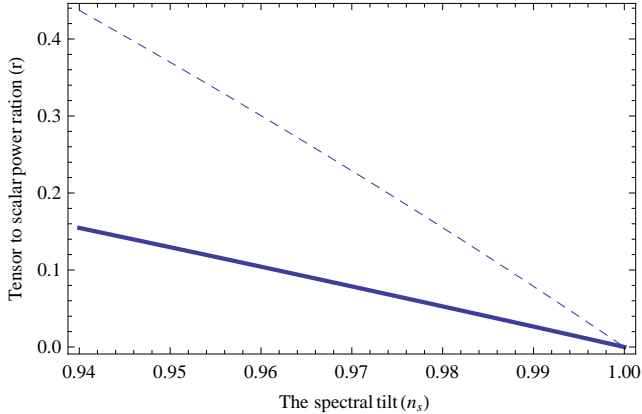}
\caption{\label{fig:epsart} The ( $n_{s}$, $r$) plane of the model. Dashed
line gives the results for $V(\varphi)=M^{6}/\varphi^{2}$. Solid line represents predictions of the other potentials.}
\end{figure}
\section{\label{sec:level1}	Summary and discussion}
In the simple theory of inflation, i.e. \eqref{simple}, two independent parameters are used to control the dynamics of the scalar field.
The parameters can be related to the first and second order derivative of the potential with respect to its argument. Therefore, a
potential can be determined via its pattern in the $( n_{s}, r)$ plane \cite{weinberg}.\\
In this paper we analyzed a model in which almost all potential have the same pattern in the $( n_{s}, r)$. In other words, we
need just one parameter to describe the behavior of the model in the inflationary phase. Actually one of the important difference between this model and other models for inflation, is appeared in the left hand side of Eq. \eqref{1-1-2}. The present of $H^{2}$ in the denominator of the expression, suppresses evaluation of the scalar field by its dynamics. Therefore, it is possible to give slow roll conditions without imposing condition on second derivative of a potential.\\
By comparing the predictions of the model in Fig.1 to the Plank data in \cite{plank-data}, we expect that the future data from the Plank satellite help
us to understand whether the model can be used for the early Universe.\\
In this work, We have focused on the early Universe just to show facility of the model. The exact solution in
Sec. II shows that how by just demand reasonable conditions on the solution, the accelerated expansion phase is arisen which itself is interesting result.       
\begin{acknowledgments}
I am grateful for helpful discussions with F. Arash.
\end{acknowledgments}
\appendix
\section{}
In \cite{higgse} The second order actions for the metric perturbations obtained.\\
The action for the scalar metric perturbation is given by
\begin{eqnarray}\label{Aaction}
S^{s} &=& \frac{1}{\kappa^{2}}\int d^4x a^3 \Bigg[ \frac{\Gamma ^2\Sigma}{H^2}{\dot \zeta}^2
-\frac{ \epsilon_s }{a^2}(\partial_i\zeta)^2 \Bigg],
\end{eqnarray}
where
\begin{equation}
\epsilon_s \equiv  \frac{d}{adt} \left[\frac{a\Gamma}{H}\left( 1-\frac{\kappa^{2}\alpha^{2}\dot{\phi}^2}{2} \right)  \right]  -\left( 1+\frac{\kappa^{2}\alpha^{2}\dot{\phi}^2}{2} \right) ,
\end{equation}
\begin{equation}
\Gamma \equiv \frac{2-\kappa^{2}\alpha^{2}\dot{\phi}^2}{2- 3\kappa^{2}\alpha^{2}\dot{\phi}^2},
\end{equation}
and
\begin{equation}
 \Sigma \equiv \frac{\kappa^{2}\dot{\phi}^2}{2}
\left[1+\frac{3\alpha^{2}H^2 (1+ \frac{ 3\kappa^{2}\alpha^{2}\dot{\phi}^2 }{ 2 }) }{(1-\frac{ \kappa^{2}\alpha^{2}\dot{\phi}^2 }{ 2 }) }\right].
\end{equation}
The action for the tensor metric perturbation is given by
\begin{eqnarray}\label{Action2}
S^{T}=\frac{1}{8\kappa^{2}}\int d^{4}x&&[(1+\frac{\alpha^{2}\kappa^{2}{\dot{\varphi}}^{2}}{2})a\gamma_{ij}\partial^{2}\gamma_{ij}\nonumber\\
&&+(1-\frac{\alpha^{2}\kappa^{2}{\dot{\varphi}}^{2}}{2})a^{3}\dot{\gamma_{ij}}\dot{\gamma_{ij}}].
\end{eqnarray}
\bibliography{apssamp}

\begin{thebibliography}{99}
\bibitem{refadd}
Actually, the procedure that we used in this paper is borrowed from Weinberg's book in Ref. [2]. Weinberg used the fact that for
the exponential potential, the simple model, (1), can be solved exactly. Then, other potentials are studied by slow-roll approximation method. Note that for the exponential potential in (1), the inflation would not end. But, as Weinberg pointed out, it is valuable to study the metric perturbations for this case. We find that this procedure is also valuable for our model.
\end{thebibliography}

\end{document}